\documentstyle[twocolumn,epsfig]{jpsj}
\newcommand{\vereq}[2]{\lower3pt\vbox{\baselineskip1.5pt \lineskip1.5pt
\ialign{$#1\hfill##\hfil$\crcr#2\crcr\sim\crcr}}}

\newcommand{\gtrsim}{\mathrel{\mathpalette\vereq>}}

\newcommand{\ltrsim}{\mathrel{\mathpalette\vereq<}}

\def\runtitle{Instability of dilute granular flow 
on rough slope}
\def\runauthor{Namiko {\sc Mitarai} and Hiizu {\sc Nakanishi}}

\title
{Instability of dilute granular flows on rough slope
}

\author
{ 
Namiko {\sc Mitarai}\footnote{E-mail: namiko@stat.phys.kyushu-u.ac.jp} 
and Hiizu {\sc Nakanishi}\footnote{E-mail: naka4scp@mbox.nc.kyushu-u.ac.jp} 
}

\inst
{
Department of Physics, Kyushu University 33, Fukuoka 812-8581
}

\recdate
{
\today
}

\abst
{
We study numerically the stability of granular flow
on a rough slope in collisional flow regime in the two-dimension.
We examine the density dependence of the flowing behavior
in low density region, and
demonstrate that the particle collisions stabilize the 
flow above a certain density in the parameter region
where a single particle shows an accelerated behavior.
Within this parameter regime, however, 
the uniform flow is only metastable
and is shown to be unstable against 
clustering when the particle density is not high enough.
}

\kword
{
granular flow, surface flow, clustering, 
inelastic collision, simulation,
discrete element method
}

\begin{document}
\sloppy
\maketitle
Granular flow on a slope is one of the simplest 
situations to see the characteristic behavior of granular dynamics.
When the inclination angle is smaller than 
a certain value (the angle of repose), the material never flows
because of the stress sustained by the friction. 
Beyond that angle, the surface layers of the material
may flow like a fluid, but the bottom part of the materials
may remain solidified when the inclination angle 
is not steep enough.~\cite{JN96}
If the inclination is increased further, all of the material
starts to flow rapidly, and 
the interaction between particles or between a particle and the slope is
dominated by the inelastic collision, rather than friction.

Many researches have been done in such a collisional flow regime, 
but most of them focus on the property of the 
uniform flow.~\cite{D91, P93, ACM99} 
In such researches, the depth dependence of the flow properties
such as velocity or density profile is investigated,
assuming that the flow is uniform in the direction
along the slope.
It is known, however, that the granular materials have the tendency 
to cluster due to the inelastic collision,
which causes the 
formation of density waves in the case of granular flow in a vertical
pipe.~\cite{HNNM95,L94} 
Therefore, it is natural to expect 
that this tendency will cause some instability 
in the uniform flow on a slope.

Granular flow of an independent particle, or a single particle 
behavior, has been studied, and 
has been found to show three types of motion 
depending on the inclination angle and roughness of the 
slope.~\cite{RJRHB94,RRB94,DBW96}
For the fixed roughness of the slope,
the following behaviors are observed upon increasing the inclination angle:
(i)The particle stops after a few collisions with the slope 
 for any initial velocity (regime A). 
(ii)The particle quickly reaches a constant averaged velocity 
in the direction along the slope and shows almost steady motion;
the averaged velocity does not depend on the initial
condition (regime B).
(iii)The particle jumps and accelerates as it goes down the slope
(regime C).

In the present work, we study how the above single particle
picture is modified in the collisional flow with 
finite density by the particle collisions.
Based on numerical simulations, we determine the parameter region
where the uniform collisional flow is realized with the finite density,
and examine the stability of the uniform flow.

\begin{figure}[t]
\begin{center}
\epsfig{width=5.5cm,file=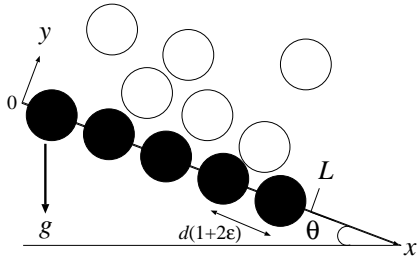}
\end{center}
\caption{A schematic illustration of the system geometry. }
\label{slope}
\end{figure}

We employed the discrete element method~\cite{CS79,FUN} with the normal and 
the tangential elastic force and dissipation.
In the simulations, granular particles are modeled by 
two-dimensional disks with mass $m$ and diameter $d$.
When the two disks $i$ and $j$ at positions $\mib{r}_{i}$ and $\mib{r}_j$ 
with velocities $\mib{v}_{i}$ and $\mib{v}_j$
and angular velocities $\omega_{i}$  and $\omega_j$ are
in contact, the force acting on the particle $i$ from the particle $j$
is calculated as follows:
The normal velocity $v_n$ and the tangential velocity $v_t$
are given by
\begin{equation}
v_n=\mib{v}_{ij}\cdot \mib{n},\quad
v_t=\mib{v}_{ij}\cdot\mib{t}-(d/2)(\omega_i+\omega_j),
\end{equation}
with the normal vector
$\mib{n}=\mib{r}_{ij}/|\mib{r}_{ij}|=(n_x,n_y)$ and
the tangential vector $\mib{t}=(-n_y,n_x)$.
Here, $\mib{r}_{ij}=\mib{r}_j-\mib{r}_i$
and $\mib{v}_{ij}=\mib{v}_j-\mib{v}_i$. 
Then the normal force $F^n_{ij}$ 
and the tangential force $F^t_{ij}$ 
acting on the particle $i$ from the particle $j$ are given by
\begin{eqnarray}
F^n_{ij}&=&-k_n(d-|\mib{r}_{ij}|)+\eta_n v_n,\\
F^t_{ij}&=&\min(|h_t|,\mu|F^n_{ij}|) \mbox{sign}(h_t),
\end{eqnarray}
with 
\begin{equation}
h_t=k_tu_t+\eta_tv_t, 
\quad 
u_t=\int^t_{t_0}v_t\mbox{d}t.
\end{equation}
Here, $u_t$ is the tangential displacement, and
$t_0$ is the time when the particles started to contact.
$k_{n}$ and $k_t$ are elastic constants,
$\eta_{n}$ and $\eta_t$ are damping parameters, 
and $\mu$ is the Coulomb friction coefficient for sliding friction.
Each particle is also subject to the gravity, and
the gravitational acceleration is given by 
$\mib{g}=g(\sin\theta,-\cos\theta)$.
The surface of the slope is made rough by gluing the particles
identical with the moving ones with spacing $2\epsilon d$ 
with $\epsilon=0.001$ (Fig. \ref{slope}).
The periodic boundary condition is imposed
in the $x$ direction.
The system size $L$ is
determined by the number of the particles glued on the slope $n_s$
as $L=(1+2\epsilon)dn_s$.
\begin{table}[t]
\caption{Parameters used in the simulations.}
\begin{center}
\begin{tabular}{ccccc}\hline
$k_n$&$k_t$&$\eta_n$&$\eta_t$&$\mu$\\
\hline
$5\times 10^4$&$(5\times 10^4)/3$&$35.68$&$35.68/3$&$0.5$\\ \hline
\end{tabular}
\label{parameter}
\end{center}
\end{table}

All quantities which appear in the following 
are given in the non-dimensionalized
form in terms of the length unit $d$, the mass unit $m$, 
and the time unit $\tau=(d/g)^{1/2}$.
The parameters used in the simulations are tabulated in Table \ref{parameter}.
With these parameters, the normal restitution coefficient 
(the ratio of normal relative velocities of pre- and post-collision) 
$e_n$ turns out to be around $0.7$.
We integrated the equations of motions using the predictor-corrector
method in the third order with a constant time step 
$\delta t=1\times 10^{-4}$.

We first investigate how the uniform collisional flow appears 
when we increase the density of flowing particles 
from the one particle limit.

As we have referred, a particle on a slope 
shows three types of behavior depending on the value of $\theta$.
It is known, however, that the boundary between the regime B 
(the constant velocity regime) and C (the acceleration regime) is not
sharp; there is a parameter region
where the particle can attain steady state or
can accelerate, depending on the initial condition.
We should also note that, when the initial kinetic energy of the particle
is too small, the particle may stop in the regime B and in 
part of the regime C.
In the present model with the parameters in Table \ref{parameter}, 
it turns out that the regime A roughly corresponds to 
the region $\sin\theta \ltrsim 0.11$, 
the regime B to $0.11\ltrsim\sin\theta\ltrsim 0.14$,
and the regime C to $0.16\ltrsim\sin\theta$.

Now we focus on what happens when the number of moving particles $n$ is
increased with the finite $L$.
It is easy to expect that the dissipation becomes 
larger upon increasing $n$, because the collisions 
between particles or a particle and the slope become
more frequent.
Therefore it is obvious that 
all the particles will stop irrespective of the initial
condition if $\theta$ is in the regime A.
Even for any $\theta$ within the regime B,
it turns out that the collisions between 
particles eventually stop all the particles.

We examine closely the flowing behavior for the larger inclination
angle $\theta$ in the regime C.
The situation is presented by showing the simulations with
$\sin\theta=0.45$ and $L=20.04$.
We set the initial condition as follows:
\begin{eqnarray}
&&x_i(0)=(L/n)(i-1),\ y_i(0)=(1+2\epsilon)d+\alpha\xi_i,\label{eq:ini1}\\
&&u_i(0)=0,\ v_i(0)=0,\ \omega_i(0)=0,\label{eq:ini2}
\end{eqnarray}
where $x_i(t)$ ($y_i(t)$) and $u_i(t)$ ($v_i(t)$) are
the coordinate and the velocity of center of mass of
the $i$th particle at time $t$ in the $x$ ($y$) direction, respectively.
$\xi_i$ is a random number uniformly distributed 
over the interval [$0$, $1$]. 
The appropriate value of $\alpha$ is a few times the diameter, 
and we adopted $\alpha=3\sim 4$ in actual simulations.
In the following, we define the 'density' of the particles as
$\rho=n/L$.
In the simulations with $L=20.04$,
we do not observe non-uniformity along the slope in the particle
distribution, 
therefore the parameter $\rho$ is enough to describe the situation.
In order to characterize the qualitative difference of 
accelerated behavior in the regime C and the uniform flow,
we investigate the $\rho$ dependence of the averaged kinetic energy
$\bar E(t)$ defined as
\begin{equation}
\bar E(t)=\frac{1}{n}\sum_{i=1}^nE_i(t),
\end{equation}
where $E_i(t)$ is the kinetic energy of the $i$th particle
at time $t$; 
\begin{equation}
E_i(t)=\frac{1}{2}m\left[u_i(t)^2+v_i(t)^2\right]+\frac{1}{2}I\omega_i(t)^2,
\end{equation}
with $I=md^2/8$.

\begin{figure}[t]
\begin{center}
\epsfig{angle=-90,width=7cm,file=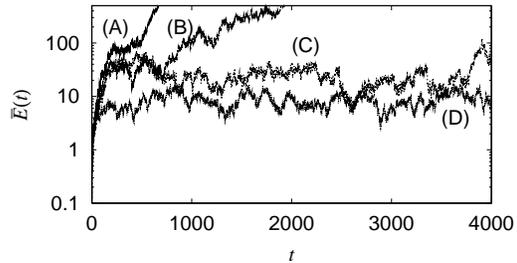}
\end{center}
\caption{The semi-log plot of $\bar E(t)$ for 
(A)$\rho=0.50$, (B)$\rho=0.60$, (C)$\rho=0.65$, and 
(D)$\rho=1.00$, with $\sin\theta=0.45$ 
and $L=20.04$. $\bar E(t)$ continues to grow when $\rho\le 0.60$, 
while it is bounded when $\rho\ge 0.65$. 
}
\label{barE}
\end{figure}
In Fig. \ref{barE}, the time evolutions of $\bar E(t)$
are shown for several values of $\rho$.
When $\rho$ is small ($\rho\le 0.60$),
$\bar E(t)$ grows rapidly and continuously, but
the growth rate become smaller as $\rho$ increases.
In this region, each particle jumps and rarely collides
with each other.
When $\rho\ge 0.65$,
$\bar E(t)$ still grows rapidly in early stage, but
its long-time behavior seems to be bounded 
and fluctuate around a constant value.
In this case each particle also jumps, but often collides with 
other particles and is prevented from jumping up infinitely.
Based on this observation, 
we define the uniform flow in the low-density limit
as the flow which is uniform along the flow direction 
with the value of $\bar E(t)$ being bounded,
namely the energy dissipation due to inelastic collisions
balances with the energy gain from the gravity.

The average value of the kinetic energy $\bar E$
in the uniform flow should be related to $X$,
the distance between collisions along the slope,
by $(1-e_n^2)\bar E\sim mgX\sin\theta$,
because the energy gain by the gravitation should balance
with the loss due to inelastic collisions.
The simulation shows the typical value of $X$
to be $O(1)\sim O(10)$ in the uniform flow,
thus we expect $\bar E$ to be $O(1)\sim O(10)$.
On the other hand, 
$\bar E$ diverges in the accelerated regime.

\begin{figure}[t]
\begin{center}
\epsfig{angle=-90,width=7cm,file=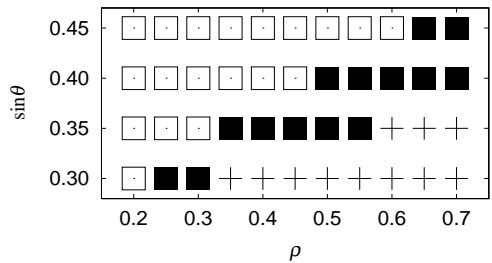}
\end{center}
\caption{
The 'phase diagram' obtained from the simulations with $L=20.04$.
The accelerated regime (open box), 
the uniform flow regime (filled box), 
and the regime where all the particles stop (plus)
are shown when more than 50 \% of trials result in the 
corresponding behavior.
}
\label{phase}
\end{figure}

Figure \ref{phase} shows the 'phase diagram' 
summarizing the behavior
in terms of $\theta$ and $\rho$ obtained from numerical simulations.
In order to obtain the diagram, 10 simulations with $L=20.04$
were done for each $\theta$ and $\rho$.
We identify the three behaviors, namely
(i) accelerated motion,
(ii) uniform flow,
and (iii) static state where all the particles come to rest,
by the following criterion; 
the system is determined to be in the accelerated motion
(in the static state) if
$\bar E(t)$ exceeds 500 (becomes zero)
by $t=20,000$.
It is determined to be the uniform flow if neither of them happen.
The final behavior sometimes depends on the initial condition.
This, we believe, is due to the finite size effect.
Each point in Fig. \ref{phase} is determined to be in one of these
regimes if more than 50 \% of the trials show
the one of the above behaviors.
It is expected that, if we could take the $L\to \infty$ limit
with fixed $\rho$
with keeping the particle distribution uniform along the slope,
such initial condition dependence should disappear.

\begin{figure}[t]
\begin{center}
\epsfig{angle=-90,width=7cm,file=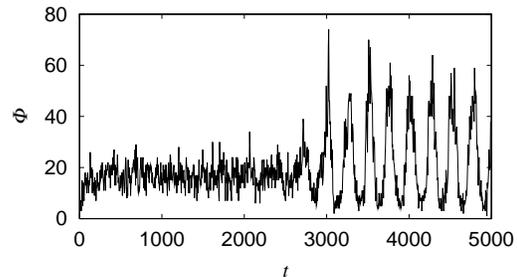}
\end{center}
\caption{
The time evolution of the flux $\Phi(t)$ 
in the simulation with $L=1002$ and $\rho=1.0$.
A fluctuation grows to form a cluster spontaneously.
}
\label{flux1000}
\end{figure}

Now we examine the stability of the uniform flow by
taking the system size large enough to observe
non-uniformity along the slope.
First we show the simulation results of the flow in the system 
with $\sin\theta=0.45$,
$L=1002$ and $n=1000$, i.e. $\rho=1.0$. 
The initial condition is also given by eq. (\ref{eq:ini2}),
therefore the particles are uniformly distributed along the slope at first.
Figure \ref{flux1000} shows the time evolution of the flux $\Phi(t)$
defined as the number of the particles which pass $x=L/2$
during the time interval $\Delta t=10$.
After a short transient time, the flux becomes almost constant  
with small fluctuations ($300\ltrsim t\ltrsim 2000$),
which indicates uniform flow is realized.
Then $\Phi(t)$ shows large fluctuation which grows in the course of time
($2000\ltrsim t \ltrsim 3000$); 
this means the clustering behavior is triggered by the fluctuation.  
Finally $\Phi(t)$ begins to oscillate almost periodically 
with large amplitude ($t\gtrsim 3000$). 
This oscillation of the flux indicates 
that one large cluster of particles 
travels in the system with almost constant velocity; 
we can see that the cluster is stable once it is formed.
From this observation, we expect that 
the uniform flow is not stable against clustering.
Considering the fact that the uniform flow has a finite life time
during which the flow seems fairly stable,
we expect that a fluctuation of finite size is 
necessary to trigger the clustering,
which means the uniform flow is metastable.
\begin{figure}[t]
\begin{center}
\epsfig{angle=-90,width=6.8cm,file=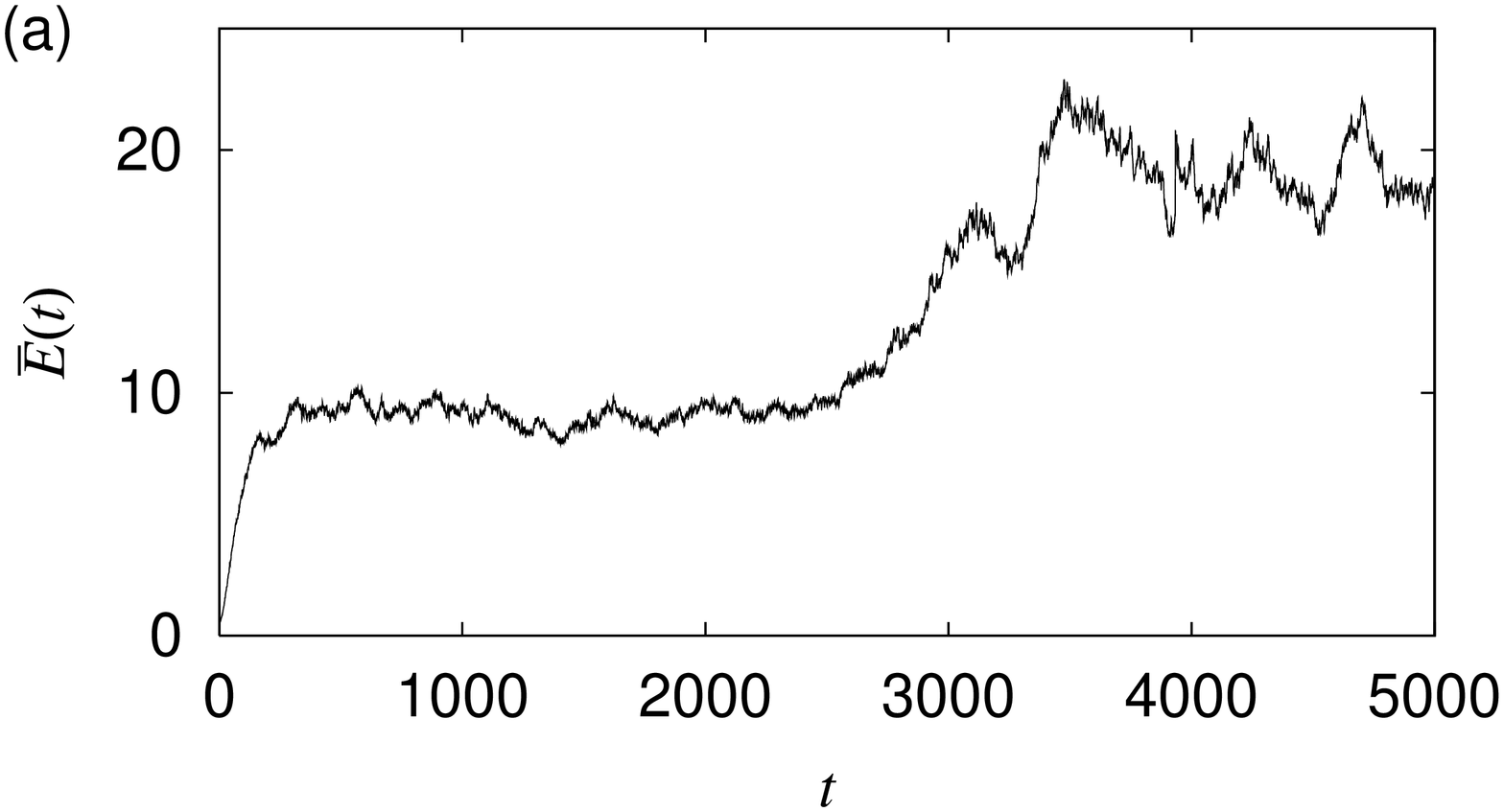}
\epsfig{angle=-90,width=6.8cm,file=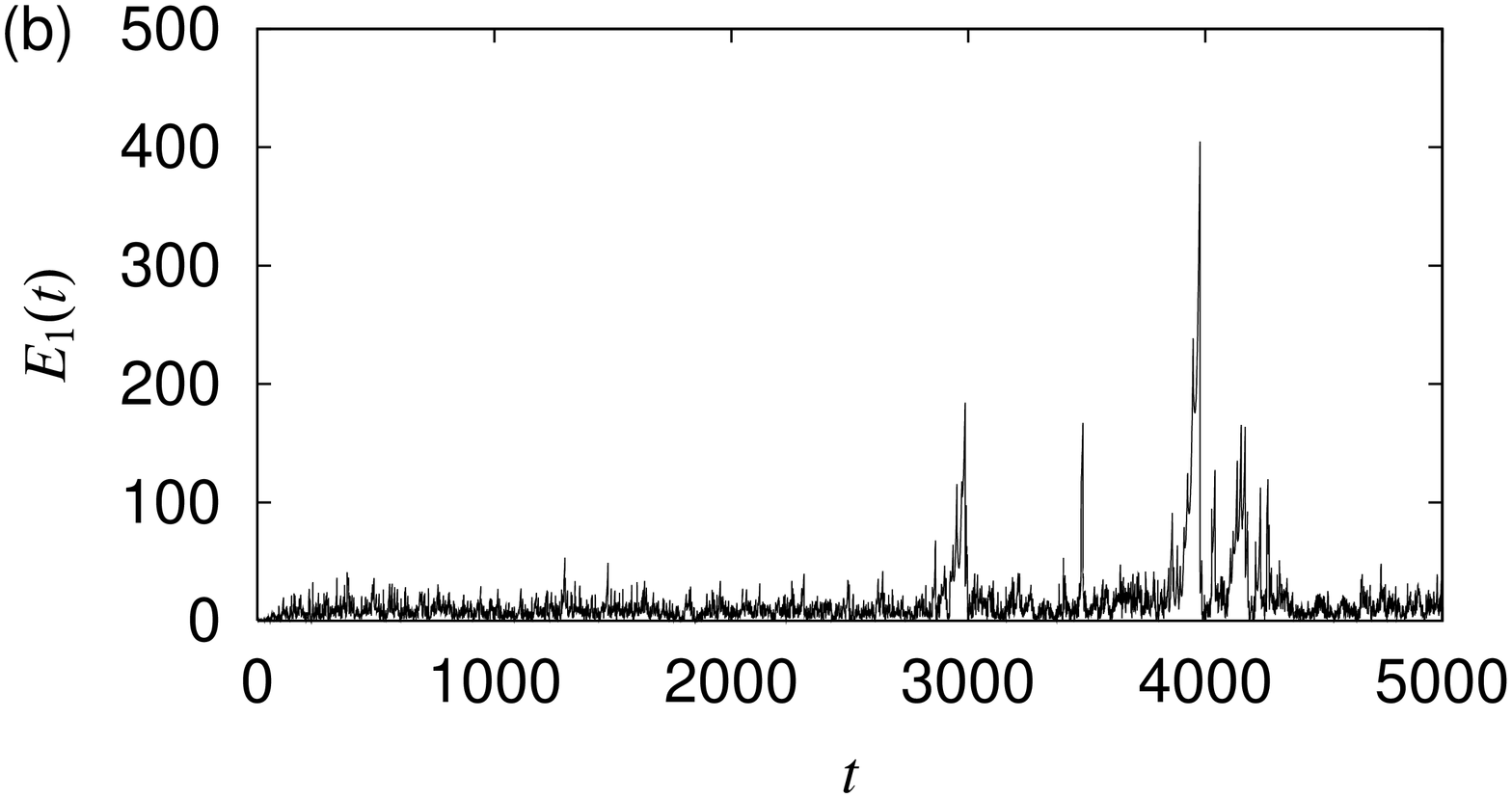}
\end{center}
\caption{
The time evolution of (a) $\bar E(t)$ and (b) $E_1(t)$
in the simulation with $L=1002$ and $\rho=1.0$.
Compared with Fig. \ref{flux1000}, we can see that,
after the clustering occurs ($t\gtrsim 3000$),
the time averaged value of $\bar E(t)$ 
becomes larger than that in the uniform flow stage,
and $E_1(t)$ begins to show large fluctuation.
}
\label{energy}
\end{figure}

In Figs. \ref{energy},
the time evolutions of (a) the averaged kinetic energy 
$\bar E(t)$ and (b) the kinetic energy of a particular particle
$E_1(t)$ are also shown.
Both of them are almost constant in the stage of the uniform flow.
Then $\bar E(t)$ increases in $2000\ltrsim t\ltrsim 3000$.
In $t\gtrsim 3000$, the fluctuation of $E_1(t)$ becomes considerably
large, and $\bar E(t)$ begins to fluctuate around another constant value which
is larger than the one in the stage of the uniform flow.
The reason why the kinetic energy increases when a cluster is 
formed can be understood as follows:
When the cluster is formed, the region with
the density lower than the threshold value 
to prevent the acceleration
($\sim 0.65$ in the case of $\sin\theta=0.45$, see Fig. \ref{phase}.)
appears locally, and particles in such a spatial region can 
be highly accelerated. 
However, the particle 
will be caught in the cluster sooner or later, 
and then quickly lose its kinetic energy.
This mechanism maintains the moving cluster, 
and results in the large fluctuations in $E_1(t)$.

\begin{figure}[t]
\begin{center}
\epsfig{angle=-90,width=7cm,file=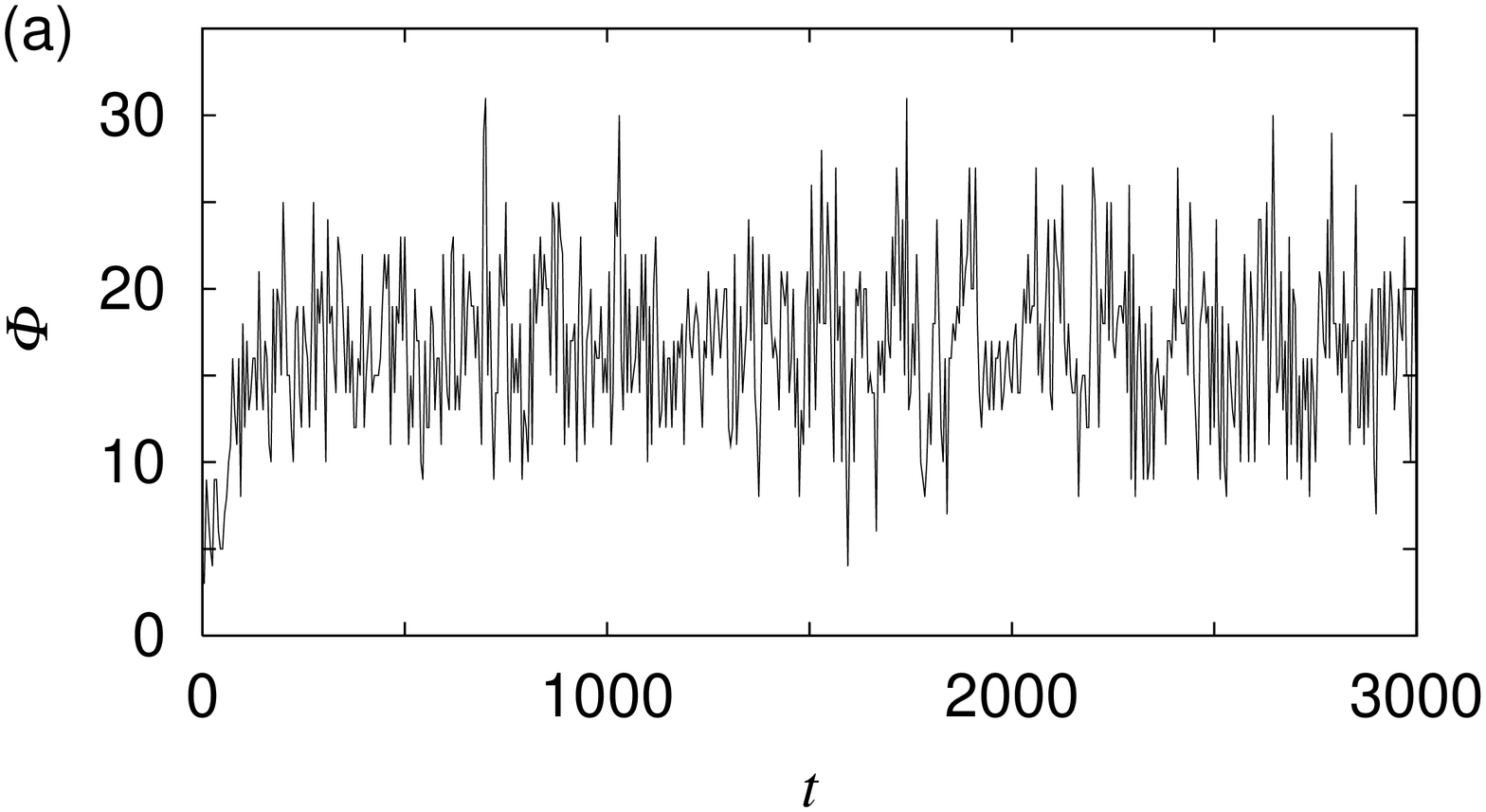}
\epsfig{angle=-90,width=7cm,file=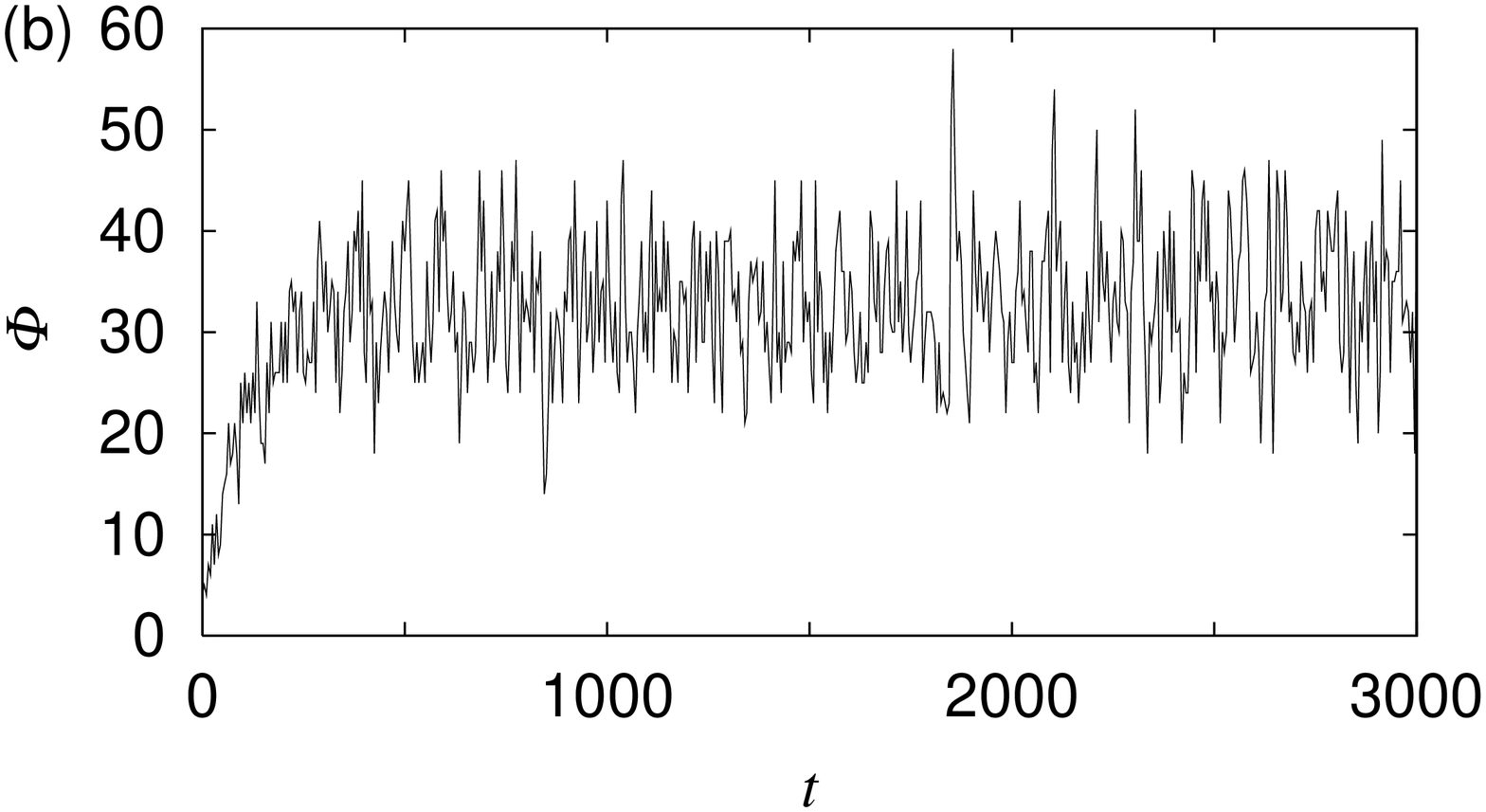}
\epsfig{angle=-90,width=7cm,file=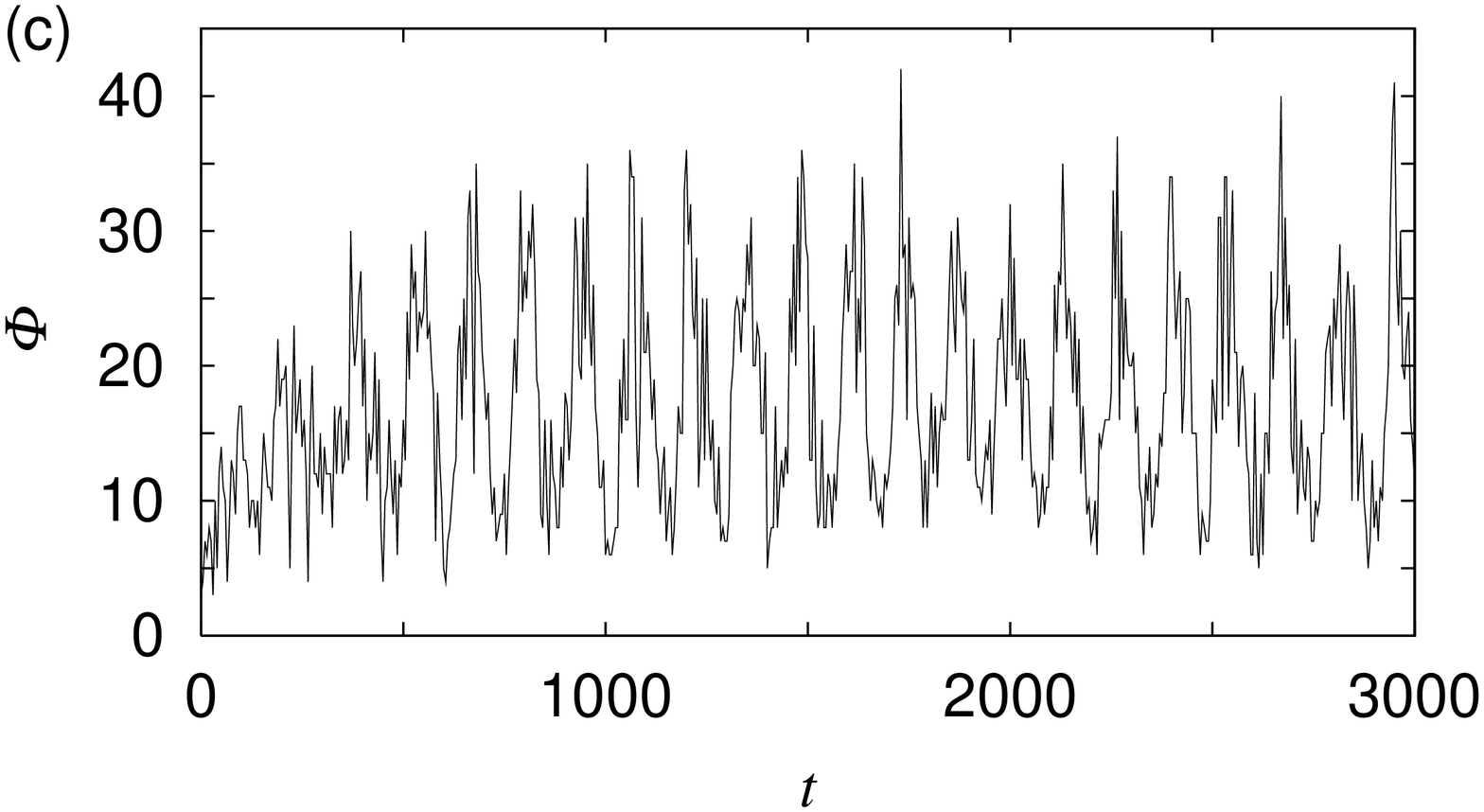}
\end{center}
\caption{
The time evolution of the flux $\Phi(t)$ 
in the simulation with $L=501$ and (a) $\rho=1.0$,
(b) $\rho=2.0$, and (c) $\rho=0.75$.}
\label{flux500}
\end{figure}
 also examined the system size and the density dependence of the 
clustering. The time evolutions of the flux for a few values of 
densities with $L=501$ are also shown in Figs. \ref{flux500}.
Figure \ref{flux500} (a) shows the result with $\rho=1.0$ ($n=500$),
which is the same density as the simulation in Fig. \ref{flux1000}.
In Fig. \ref{flux500} (a),
however, no clustering behavior can be seen clearly 
even though the fluctuation is very large.
We have also simulated in the system with $L=250.5$ and $\rho=1.0$
($n=250$),
but the clustering behavior was not found, either.
On the other hand, 
the density dependence of the system behavior
can be seen in
Figs. \ref{flux500} (b) ($\rho=2.0$)
and (c) ($\rho=0.75$).
It is found that the uniform flow is maintained in 
Fig. \ref{flux500} (b), while the clear oscillation of the flux
associated with the cluster formation is seen in Fig. \ref{flux500} (c).
From these results, we can see the general tendency that 
the uniform flow tends to be stabilized 
as the system size is smaller or as the density is higher. 

In summary, we have examined the two-dimensional 
granular flow on a rough slope
in the collisional flow regime by numerical simulations.
It was shown that the mutual collisions among particles
stabilizes the flow even in the accelerated regime 
for a single particle system.
The phase diagram was determined for the accelerated, uniform flow,
and stopping regime in terms of the particle density $\rho$
and the inclination angle $\theta$
for a particular system size.
The stability of uniform flow was also examined and we
found that a large single cluster appears spontaneously
out of a uniform initial state.
It was shown that the smaller the particle density is,
or the larger the system size is,
the less stable the uniform flow is.

Before concluding, let us make a comment on the
system size dependence of the instability.
Consider the situation where one large cluster,
whose length along the slope is $l$ and
typical hight is $h$, is formed. 
If we can neglect the particles
outside the cluster,
we can assume, as the first approximation, that the volume of 
the cluster ($\propto lh$ for 2-dimensional case)
is proportional to $n$. 
Namely we get the relation $(l/d)(h/d)\propto n =(\rho d)(L/d)$.
Therefore, if $(l/d)$ and $(h/d)$ are scaled with $(L/d)$ for fixed $\rho$
as $(l/d)\propto (L/d)^\beta$ and $(h/d)\propto (L/d)^{\beta'}$,
we get the relation $\beta'=1-\beta$.
Because both $(l/d)$ and $(h/d)$ should be increasing function of
$(L/d)$ for fixed $\rho$, we get an inequality for the exponent $\beta$
that $0<\beta<1$.
Then there should exist an critical system size $L_c$ such that
$l(L_c)\sim L_c$;
a cluster can exist only in the case of $L>L_c$ for which $l(L)<L$,
and the uniform flow become stable when $L<L_c$.
Using similar argument,  
we can also predict the existence of 
the critical density $\rho_c$ for a fixed $L$.
However, we should note that the moving cluster
seems to be maintained by the balance 
between the outgoing flux of the particles at the front 
of the cluster and incoming flux at the tail,
therefore the motion of the particles outside the
cluster may affect on the shape of the cluster in
the periodic boundary condition.
Therefore, the discussion here may not simply hold
and the exponent $\beta$ may depend on $\rho$.
The scaling behaviors should be confirmed more carefully.

The spontaneous cluster formation 
out of uniform flow is seen also 
in the granular flow through a vertical 
pipe~\cite{L94,HNNM95} and 
in the traffic flow on a freeway.~\cite{CSS00,MN00}
It is an interesting problem 
to find out whether those phenomena and 
the cluster formation in surface flow
have a common mathematical structure
at phenomenological level.

Part of the computation in this work has been done 
using the facilities of the Supercomputer Center, 
Institute for Solid State Physics, University of Tokyo. 


\end{document}